\documentclass[aps,prb,floatfix,nofootinbib,superscriptaddress]{revtex4-2}
\usepackage{graphicx}
\usepackage{amsmath,amssymb,mathrsfs,amsthm}
\usepackage[pdftex, colorlinks=true, linkcolor=blue, citecolor=blue, urlcolor=blue]{hyperref}
\usepackage{bbold}
\usepackage{mathtools}
\usepackage{braket}
\usepackage{comment}

\usepackage{lipsum}
\usepackage[normalem]{ulem}
\usepackage{units}
\usepackage{color}
\usepackage{float}

\let\isout\sout
\renewcommand{\sout}[1]{\ifmmode\text{\isout{\ensuremath{#1}}}\else\isout{#1}\fi}

\begin{document}

\title{Reentrant localization transition in a dimerized quasiperiodic dipolar chain}

\author{Thomas F.\ Allard}\email{thomas.allard@uam.es} 
\affiliation{Universit\'e de Strasbourg, CNRS, Institut de Physique et Chimie des Mat\'eriaux de Strasbourg, UMR 7504, F-67000 Strasbourg, France}
\affiliation{Departamento de F\'{i}sica Te\'{o}rica de la Materia Condensada, Universidad Aut\'{o}noma de Madrid, E-28049 Madrid, Spain}
\affiliation{Condensed Matter Physics Center (IFIMAC), Universidad Aut\'{o}noma de Madrid, E-28049 Madrid, Spain}

\author{Guillaume Weick}
\affiliation{Universit\'e de Strasbourg, CNRS, Institut de Physique et Chimie des Mat\'eriaux de Strasbourg, UMR 7504, F-67000 Strasbourg, France}

\begin{abstract}
Reentrant localization transitions, that is, the transitions of a portion of the eigenspectrum from localized to critical and then again to localized as the quasiperiodic modulation strength is increased, have been recently unveiled in various quasiperiodic models.
However, both the physical mechanisms underlying these transitions and how they may extend to systems with long-range coupling and dissipation remain elusive.
Here we investigate the fate of such a phenomenon in a dimerized quasiperiodic chain of lossy dipolar emitters with all-to-all coupling.
We demonstrate that in this model, reentrant transitions survive to all-to-all couplings and occur from an interplay between the chain dimerization and an asymmetric quasiperiodic modulation of the emitter spacings.
Transport simulations through a driven-dissipative open quantum system approach complete our study and reveal the detrimental effects of emitter losses on the reentrant localization transition.
\end{abstract}

\maketitle

%%%%%%%%%%%%%%%%%%%%%%%%%%%%%%%%%%%%%%%%%%%%%%%%%%%%%%%%%%%%%%%%%%%%%%%%%%%%%%%%%%%%%%%%
%%%%%%%%%%%%%%%%%%%%%%%%%%%%%%%%%%%%%%%%%%%%%%%%%%%%%%%%%%%%%%%%%%%%%%%%%%%%%%%%%%%%%%%%
%%%%%%%%%%%%%%%%%%%%%%%%%%%%%%%%%%%%%%%%%%%%%%%%%%%%%%%%%%%%%%%%%%%%%%%%%%%%%%%%%%%%%%%%
\section{Introduction}
\label{sec:Introduction}
%%%%%%%%%%%%%%%%%%%%%%%%%%%%%%%%%%%%%%%%%%%%%%%%%%%%%%%%%%%%%%%%%%%%%%%%%%%%%%%%%%%%%%%%
%%%%%%%%%%%%%%%%%%%%%%%%%%%%%%%%%%%%%%%%%%%%%%%%%%%%%%%%%%%%%%%%%%%%%%%%%%%%%%%%%%%%%%%%
%%%%%%%%%%%%%%%%%%%%%%%%%%%%%%%%%%%%%%%%%%%%%%%%%%%%%%%%%%%%%%%%%%%%%%%%%%%%%%%%%%%%%%%%

The Aubry-André (AA) model of a one-dimensional (1d) quasiperiodic crystal consists of a lattice with incommensurably modulated onsite energies and constant nearest-neighbor couplings.
This tight-binding model enables one to explore the Anderson transition \cite{Aubry1980} as well as the hopping of electrons on a two-dimensional lattice under a magnetic field \cite{Harper1955}. The AA model is since the past few decades an entire field of study by itself \cite{Domínguez-Castro_2019}, drawing the attention of both the physical \cite{Harper1955,Aubry1980,Domínguez-Castro_2019} and mathematical \cite{Jitomirskaya1999,Avila2009} communities.
On the experimental side, pioneering realizations of the AA model have been achieved using cold atoms in a superposition of two optical lattices with incommensurate wavelengths \cite{Roati2008}, photonic lattices of evanescently coupled waveguides with quasiperiodically modulated widths \cite{Lahini2009}, and more recently in plasmonic gold nanochains \cite{Yan2025}.

Many extensions of the AA model with specific forms of onsite energy modulation \cite{Grempel1982,Ganeshan2015}, quasiperiodically modulated couplings \cite{Han1994,Chang1997,Kraus2012,Li2023} (also termed off-diagonal modulation), dimerized chains with the latter modulation \cite{Liu2015,Cestari2016,Liu2018,Longhi2020,Xiao2021,LiuTong2022,Antao2024,Liu2016_arxiv}, or also with richer physics such as interactions \cite{Iyer2013}, power-law and long-range couplings \cite{Deng2019, Dominguez-Castro2022,Liu2024}, or non-Hermitian effects \cite{Jazaeri2001,Yuce2014,Jiang2019,Longhi2019} have then been investigated.
Recently, quasiperiodic \emph{dipolar} chains have attracted particular attention, as they constitute a versatile platform offering the opportunity to explore at the same time the effects of long-range power-law couplings, non-Hermiticity, and off-diagonal quasiperiodic modulation \cite{Wang2021, Hu2024, Citrin2024}.
Dipolar fermions \cite{Molignini2025_fermions} and bosons \cite{Molignini2025_bosons} in a quasiperiodic potential have also lately  been investigated.

Among the many interesting properties of quasiperiodic chains, a recent surge of theoretical studies 
\cite{Ramakumar2015,Goblot2020, Roy2021, Jiang2021, Wu2021, Han2022, Padhan2022, Roy2022, Wang2023, Qi2023, Guan2023, Dai2023, Goncalves2023, Goncalves2023, Shimasaki2024, Padhan2024, Ganguly2024, Guo2024, Miranda2024, Tabanelli2024, Lu2025, Nair2025, Li2023_arxiv2, Li2023_arxiv,Chang2025} 
 has been dedicated to what is known as reentrant localization transitions (RLTs). These correspond to the unusual transition of already localized eigenstates to critical and then again to localized when increasing the quasiperiodic modulation strength.
This counterintuitive additional transition goes against the usual prediction of Anderson localization.
Indeed, quasiperiodic modulation is known to emulate random disorder so that eigenstates usually remain localized after the localization transition \cite{Mirlin_RevModPhys}.

While still rather poorly understood, RLTs have been studied in various quasiperiodic models where an additional parameter to the modulation strength is present, e.g., an additional potential with alternating sign \cite{Ramakumar2015}, a continuous interpolation to the Fibonacci model \cite{Goblot2020}, or a dimerization of the chain with staggered onsite \cite{Roy2021,Jiang2021,Roy2022,Han2022,Wang2023,Nair2025,Chang2025} or off-diagonal \cite{Miranda2024} modulation.
Moreover, recent experimental works managed to observe RLTs in 1d photonic crystals of quasiperiodic thicknesses \cite{Vaidya2023} as well as in superconducting circuits with tunable qubits and couplers \cite{Li_PRR2024}.
We note that very similar reentrant transitions have also been proposed and experimentally observed in random dimer systems \cite{Zuo2022,Xu2024}.

However, although found in many systems, RLTs seem to be fragile to perturbations. Indeed, non-Hermiticities were notably found to prevent the transition \cite{Jiang2021}.
On the other hand, the impact of long-range coupling proved nontrivial as second-nearest-neighbor couplings were found to prevent the transition \cite{Wang2023}, but third-nearest-neighbor ones, which preserve the sublattice symmetry of the system, did not compete against RLTs \cite{Chang2025}.

\begin{figure}[t]
 \includegraphics[width=.5\linewidth]{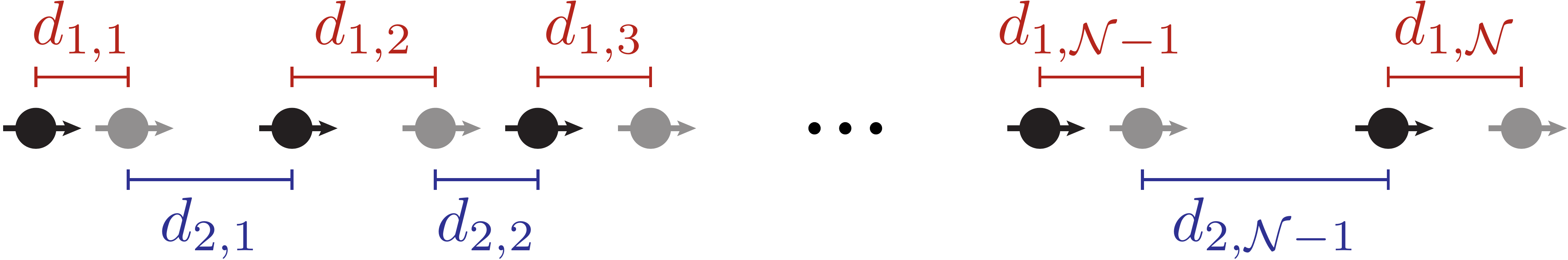}
 \caption{Sketch of the dimerized quasiperiodic dipolar chain under consideration.
 The black and grey dots with arrows represent dipolar emitters polarized longitudinally to the chain and located on the $A$ and $B$ sublattices, respectively. 
 The intra- and interdimer distances $d_{1,m}$ and $d_{2,m}$, defined in Eq.~\eqref{eq:ds}, are modulated quasiperiodically.}
 \label{fig:sketch}
\end{figure}

Motivated by these prior studies, in this work we explore RLTs in a new framework, namely, a dimerized dipolar chain with quasiperiodic modulation of the spacings between the dipolar emitters (see Fig.~\ref{fig:sketch}).
We demonstrate that in such a system, an interplay between the chain dimerization and an asymmetric quasiperiodic modulation of the spacings induces an RLT. This transition resists to the all-to-all power-law dipolar coupling that decays as one over the distance cubed, despite the detrimental effect of the latter.
Through dissipative transport simulations, we show that the aforementioned RLT is strongly affected by dissipation, but may survive to low-loss emitters.
By investigating the fate of RLTs in a dipolar chain, our study constitutes a first step to the understanding of anomalous localization transitions in more complex and realistic systems, with possible long-range couplings and inherent losses.

Our paper is organized as follows:
We present our model of a dimerized quasiperiodic dipolar chain in Sec.~\ref{sec:Model}. 
We study its localization properties in Sec.~\ref{sec:Reentrant localization transition} and investigate its transport characteristics when considering lossy emitters in Sec.~\ref{sec:Transport}.
Eventually, we draw conclusions in Sec.~\ref{sec:Conclusion}. 
We provide a detailed description of our model in Appendix~\ref{sec:app:Second quantized Hamiltonian}. 
We proceed to a multifractal analysis of the RLT in Appendix~\ref{sec:app:Multifractal analysis} and compare our results to the case of a nearest-neighbor chain in Appendix~\ref{sec:app:Nearest-neighbor}.

%%%%%%%%%%%%%%%%%%%%%%%%%%%%%%%%%%%%%%%%%%%%%%%%%%%%%%%%%%%%%%%%%%%%%%%%%%%%%%%%%%%%%%%%
%%%%%%%%%%%%%%%%%%%%%%%%%%%%%%%%%%%%%%%%%%%%%%%%%%%%%%%%%%%%%%%%%%%%%%%%%%%%%%%%%%%%%%%%
%%%%%%%%%%%%%%%%%%%%%%%%%%%%%%%%%%%%%%%%%%%%%%%%%%%%%%%%%%%%%%%%%%%%%%%%%%%%%%%%%%%%%%%%
\section{Dimerized quasiperiodic dipolar chain}
\label{sec:Model}
%%%%%%%%%%%%%%%%%%%%%%%%%%%%%%%%%%%%%%%%%%%%%%%%%%%%%%%%%%%%%%%%%%%%%%%%%%%%%%%%%%%%%%%%
%%%%%%%%%%%%%%%%%%%%%%%%%%%%%%%%%%%%%%%%%%%%%%%%%%%%%%%%%%%%%%%%%%%%%%%%%%%%%%%%%%%%%%%%
%%%%%%%%%%%%%%%%%%%%%%%%%%%%%%%%%%%%%%%%%%%%%%%%%%%%%%%%%%%%%%%%%%%%%%%%%%%%%%%%%%%%%%%%
\subsection{Model}
\label{sec:Model:Model}

The dimerized quasiperiodic chain under study is composed of $2\mathcal{N}$ subwavelength dipolar emitters arranged in the $z$ direction (see Fig.~\ref{fig:sketch}).
We treat these emitters as generic point dipoles, which behave as classical oscillating dipoles and may model various experimental platforms, from macroscopic microwave antennas \cite{Mann2018} to microscopic magnonic spheres \cite{Pirmoradian2018}, nanoscopic plasmonic \cite{Mueller2020}, dielectric \cite{Slobozhanyuk2015} or SiC \cite{Wang2018b} particles, as well as cold atoms \cite{Browaeys2016}.
Each of these dipoles is longitudinally polarized along the $z$ direction, has an effective mass $M$, an effective charge $-Q$, and a typical length scale $a$.
Its sole dynamical degree of freedom is its displacement vector $\mathbf{h}=h\hat{z}$, which leads to an electric dipole moment $\mathbf{p} = -Q \mathbf{h}$ that oscillates at a resonance frequency $\omega_0 =  \sqrt{Q^2/Ma^3}$.

After a typical second quantization scheme \cite{Downing2017_Topological}, the Hamiltonian of a dimerized quasiperiodic chain of such dipolar emitters reads
\begin{equation}
    H = \hbar\omega_0 \sum_{m=1}^\mathcal{N} 
    \left(a_{m}^{\dagger} a_{m}^{\phantom{\dagger}}+b_{m}^{\dagger} b_{m}^{\phantom{\dagger}}\right)
   +  \frac{\hbar}{2} \sum_{\substack{m,m'=1\\(m\neq m')}}^{\mathcal{N}} \left(
    \Omega^{AA}_{m, m^\prime}  a_{m^{}}^{\dagger} a_{m^\prime}^{\phantom{\dagger}} + \Omega^{BB}_{m, m^\prime} b_{m^{}}^{\dagger} b_{m^\prime}^{\phantom{\dagger}} + \mathrm{H.c.} \right)
   + \sum_{m,m'=1}^{\mathcal{N}} \hbar \Omega^{AB}_{m, m^\prime} \left( a_{m^{}}^{\dagger} b_{m^\prime}^{\phantom{\dagger}} + \mathrm{H.c.} \right),
\label{eq:H}
\end{equation}
where the rotating wave approximation has been applied, the Coulomb gauge considered, and only the quasistatic part of the Coulomb interaction retained.
We specifically choose longitudinally polarized dipoles in order to minimize the retardation effects of the Coulomb interaction, so that we can only consider its quasistatic part \cite{Allard2021}.
In Eq.~\eqref{eq:H}, the bosonic ladder operators $a_m^\dagger$ ($b_m^\dagger$) and $a_m$ ($b_m$) respectively create and annihilate a dipolar excitation in the dimer $m \in [1,\mathcal{N}]$ and sublattice $A$ ($B$).
The quasistatic dipolar coupling strength between the emitters in the Hamiltonian \eqref{eq:H} reads
\begin{equation}
    \Omega^{s, s^\prime}_{m,m^\prime} = -\omega_0 \left( \frac{a}{r^{s,s^\prime}_{m,m^\prime}}\right)^3.
\label{eq:Omega}
\end{equation}
It decays with the inverse cube of the distance $r^{s,s^\prime}_{m,m^\prime}$ between two emitters in dimers $m$ and $m^\prime$ which are, respectively, in the sublattices $s$ and $s^\prime$.
A detailed derivation of Eq.~\eqref{eq:H} is provided in Appendix~\ref{sec:app:Second quantized Hamiltonian}.

Importantly, the quasiperiodicity in our model stems from the emitter positions, so that it enters in the coupling strength \eqref{eq:Omega} through the distances 
\begin{subequations}
\label{eq:distances}
    \begin{equation}
        r^{AA}_{m,m^\prime} = \sum_{l=\min(m,m^\prime)}^{\max(m,m^\prime) - 1} \left(d_{1,l} + d_{2,l}\right),
    \end{equation}
    \begin{equation}
        r^{BB}_{m,m^\prime} = \sum_{l=\min(m,m^\prime)}^{\max(m,m^\prime) - 1} \left(d_{1,l+1} + d_{2,l}\right),
    \end{equation}
    and
    \begin{equation}
        r^{AB}_{m,m^\prime} = \sum_{l=\min(m,m^\prime)}^{\max(m,m^\prime) - 1} \left(d_{1,l} + d_{2,l}\right) - \mathrm{sgn}(m-m^\prime) d_{1,m^\prime}.
    \end{equation}
\end{subequations}
Here the intradimer spacing between an emitter $m$ on the sublattice $A$ and its neighbor to the right (see Fig.~\ref{fig:sketch}) reads
\begin{subequations}
\label{eq:ds}
    \begin{equation}
        d_{1,m} = d_1 \left[ 1 + \Delta_1 \cos\left(2\pi m \beta + \phi\right) \right], \,\,\, m \in [1,\mathcal{N}],
    \label{eq:d1}
    \end{equation}
    while the interdimer one between an emitter $m$ on the sublattice $B$ and its neighbor to the right reads
    \begin{equation}
        d_{2,m} = d_2 \left[ 1 + \Delta_2 \cos\left(2\pi m \beta + \phi\right) \right], \,\,\, m \in [1,\mathcal{N}-1].
    \label{eq:d2}
    \end{equation}
\end{subequations}
The parameter $\beta$, chosen irrational so that the distances never exactly repeat themselves along the chain, determines the incommensurate period of the quasiperiodic modulation and $\phi$ is a given phase.
As this work focuses on the localization properties, the latter phase is inconsequential to the results presented, so that we consider in the sequel 
$\phi=0$. We note that for $\phi\neq0$, such off-diagonal quasiperiodic systems feature peculiar topological properties on their own \cite{Kraus2012}, which are out of the scope of the present work.\footnote{While we will focus here on the localization properties of the bulk system, we note that such a dimerized chain with off-diagonal (quasi-)disorder is also of particular interest to study its topological properties \cite{Antao2024}, as it is notably known to host topological Anderson insulator phases, i.e., disorder-induced topological edge states \cite{Meier2018,Longhi2020,Ren2024}.}

In the distribution of spacings \eqref{eq:ds}, $\Delta_1$ and $\Delta_2$ are the quasiperiodic strengths on the intra- and interdipole distances $d_{1,m}$ and $d_{2,m}$, which are respectively centered around the values $d_1$ and $d_2$.
In order for multipolar terms to be negligible in the Hamiltonian \eqref{eq:H}, the values of the latter quasiperiodic strengths are constrained by the condition $d_{1m}, d_{2m} \gtrsim 3a$ \cite{Park2004}.
For fixed values of $d_1$ and $d_2$, our precise modeling of dipolar emitters therefore restricts the parameter space to values of $\Delta_1$ and $\Delta_2$ such that
\begin{subequations}
\label{eq:dipolar constraint}
\begin{equation}
     \Delta_{1} \leqslant 1-3\frac{a}{d_{1}}  
\end{equation}
and    
\begin{equation}
\Delta_{2} \leqslant 1-3\frac{a}{d_{2}}.
\end{equation}
\end{subequations}
To still allow for large values of quasiperiodic strengths, we fix in the remaining the sum of the averaged spacings $d = d_1 + d_2$ to $d = 15a$.

We note that when considering nearest-neighbor coupling only, the Hamiltonian \eqref{eq:H} reduces, for small quasiperiodic modulation strength, to several already studied models.
Indeed, with $d_1 \neq d_2$ and $\Delta_2=0$, it reduces to a fully asymmetric off-diagonal dimerized AA model, i.e., a Su-Schrieffer-Heeger (SSH) model with only one type of bond modulated.
Anomalous mobility edges separating critical energy intervals from localized ones have been found in this model \cite{LiuTong2022}.
Its disordered counterpart, with random disorder instead of quasiperiodic modulation, has been thoroughly studied for its topological Anderson insulating phase \cite{Meier2018,Ren2024}.
With $d_1=d_2$ and $\Delta_1=\Delta_2$, the Hamiltonian \eqref{eq:H} approximates to the off-diagonal AA model, known to exhibit a critical phase instead of a localized one \cite{Han1994,Chang1997,Kraus2012,Li2023}.
If $d_1 \neq d_2$ but $\Delta_1=\Delta_2$, Eq.~\eqref{eq:H} approximates to a dimerized (or SSH) off-diagonal AA model \cite{Liu2015,Cestari2016,Liu2018,Longhi2020,Xiao2021,LiuTong2022,Antao2024,Liu2016_arxiv}, or equivalently to an AA model with both commensurate and incommensurate off-diagonal modulations \cite{Ganeshan2013}.
Such an SSH-AA model has recently been realized experimentally \cite{Xiao2021,Li2023,Li_PRR2024}.
Finally, considering the full all-to-all dipolar coupling as we do in the sequel, and fixing $d_1=d_2$ and $\Delta_1=\Delta_2$, the Hamiltonian \eqref{eq:H} exactly reduces to the quasiperiodic dipolar chain studied in Refs.~\cite{Wang2021,Hu2024}, where peculiar topological properties as well as a large intermediate phase have been unveiled.
In the present study, we will consider an all-to-all dipolar coupling as well as $d_1 \neq d_2$ and $\Delta_1 \neq \Delta_2$.

%%%%%%%%%%%%%%%%%%%%%%%%%%%%%%%%%%%%%%%%%%%%%%%
\subsection{Asymmetric modulation}
\label{sec:asymmetric modulation}
%%%%%%%%%%%%%%%%%%%%%%%%%%%%%%%%%%%%%%%%%%%%%%%

\begin{figure}[t]
 \includegraphics[width=.5\linewidth]{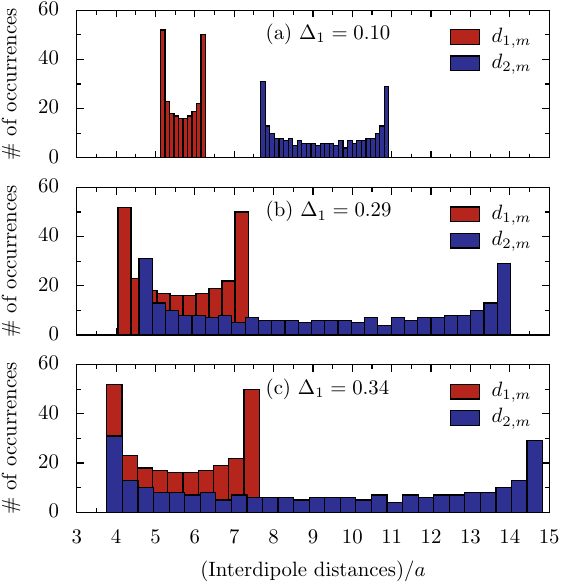}
 \caption{Histograms representing the statistical frequency distributions of the interdipole distances $d_{1,m}$ and $d_{2,m}$ (i.e., the number of occurrences within a given bin of distances), highlighting the fact that both their means and widths are unequal.
 The quasiperiodic modulation strength $\Delta_1$ is increased from (a) $\Delta_1=0.10$ to (b) $\Delta_1=0.29$ and (c) $\Delta_1=0.34$.
 All the distances are drawn from a chain composed of $\mathcal{N}=250$ dimers, with the parameters $d=15a$, $\epsilon=-0.24$, $\Gamma=1.75$, and $\beta=(\sqrt{13}+3)/2 + (\sqrt{5} + 1)/2$, specific values whose relevance will be made clear in the manuscript.}
 \label{fig:spacing distributions}
\end{figure}

The consideration of both unequal intra- and interdimer distances and unequal modulation strengths implies that the dimerization of the chain is controlled by two distinct parameters.
First, the typical (i.e., nonmodulated) dimerization 
\begin{equation}
    \epsilon = \frac{d_1 - d_2}{d},
\label{eq:epsilon}
\end{equation}
which dimerizes the chain even without any quasiperiodic modulation.
Second, the quasiperiodic strength ratio
\begin{equation}
    \Gamma = \frac{\Delta_2}{\Delta_1},
\label{eq:Gamma}
\end{equation}
which, if not unity, induces an asymmetric modulation of the intra- and intercell spacings.
This asymmetric modulation is the crucial physical ingredient of our system.
It allows the possibility of a large overlap between the intra- and interdimer spacing distributions, as one widens more than the other.
This is exemplified in Fig.~\ref{fig:spacing distributions}, in which we represent the 
frequency
distributions of $d_{1,m}$ and $d_{2,m}$ as histograms for a case with $\epsilon=-0.24$, a fixed ratio $\Gamma=1.75$, and increasing modulation strengths $\Delta_1$.
The cosine form of the modulation [see Eq.~\eqref{eq:ds}] introduces typical U-shaped arcsine distributions of distances, where the lower-end and upper-end peaks correspond to minima and maxima of distances that quasi-repeat themselves along the chain.
For $\Delta_1$ large enough, we observe in Fig.~\ref{fig:spacing distributions}(c) that the distribution of $d_{2,m}$ approximately encompasses the one of $d_{1,m}$.
Notably, the lower-end peaks are getting closer to one another, implying a significant number of dimers in which the intra- and interdimer distances become very close, effectively undimerizing parts of the chain.

\begin{figure}[t]
 \includegraphics[width=.5\linewidth]{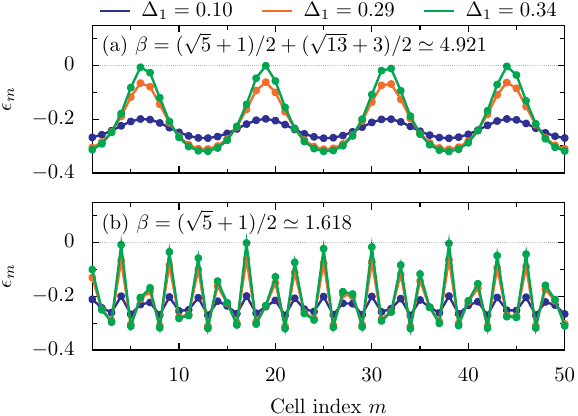}
 \caption{Local dimerization \eqref{eq:local dimerization} along the first $50$ unit cells of a quasiperiodically modulated chain with the same parameters as in Fig.~\ref{fig:spacing distributions}, considering an incommensurate period determined by (a) the sum of the bronze and golden ratios and (b) the golden ratio.}
 \label{fig:local epsilon}
\end{figure}

Indeed, an asymmetric modulation with $\Gamma\neq1$ renders the dimerization along the chain inhomogeneous, so that we can define a local, dimer-dependent dimerization as
\begin{equation}
    \epsilon_m = \frac{d_{1,m} - d_{2,m}}{d_{1,m}+d_{2,m}}, \quad m \in [1,\mathcal{N}-1].
\label{eq:local dimerization}
\end{equation}
To maximize the above-mentioned modulation-induced effective local undimerization, one needs $|\epsilon_m|\ll1$ for neighboring dimers $m$.
This leads us to consider an inverse incommensurate period $\beta$ whose fractional part is very close to $0$ or $1$, so that the term $\mathrm{cos}(2\pi m \beta)$ in Eq.~\eqref{eq:ds} changes slowly from one dimer to another.
We show the quantity \eqref{eq:local dimerization} along the first $50$ unit cells of a chain in Fig.~\ref{fig:local epsilon}.
In Fig.~\ref{fig:local epsilon}(a), we considered the same parameters as in Fig.~\ref{fig:spacing distributions}, with $\beta=(\sqrt{5}+1)/2 + (\sqrt{13}+3)/2 \simeq 4.921$ as the sum of the golden and bronze ratio.
We compare this value to the case of the golden ratio $\beta=(\sqrt{5}+1)/2\simeq1.618$ in Fig.~\ref{fig:local epsilon}(b).
While both $\beta$ are irrational and define strictly quasiperiodic sequences of distances, the fractional part close to $1$ of the sum of the bronze and golden ratio allows a slower modulation.
This leads the local dimerization to approach zero for several neighboring dimers, creating blocks of almost equidistant dipoles along the chain.
In the remaining, we therefore fix $\beta$ to the sum of the bronze and golden ratio,\footnote{We have further tested the 10th metallic mean $\beta=5+\sqrt{26}\simeq10.099$, whose fractional part is close to 0, and for which we also observed an RLT.} and study how this particular asymmetric modulation of a dimerized dipolar chain may induce unusual localization transitions.
We note that in all of our numerical simulations, we use a rational approximate of $\beta$ up to $14$ decimals. Interestingly, we have checked that the same qualitative results are found up to a truncation of $\beta$ with only $3$ decimals.

%%%%%%%%%%%%%%%%%%%%%%%%%%%%%%%%%%%%%%%%%%%%%%%%%%%%%%%%%%%%%%%%%%%%%%%%%%%%%%%%%%%%%%%%
%%%%%%%%%%%%%%%%%%%%%%%%%%%%%%%%%%%%%%%%%%%%%%%%%%%%%%%%%%%%%%%%%%%%%%%%%%%%%%%%%%%%%%%%
%%%%%%%%%%%%%%%%%%%%%%%%%%%%%%%%%%%%%%%%%%%%%%%%%%%%%%%%%%%%%%%%%%%%%%%%%%%%%%%%%%%%%%%%
\section{Localization properties}
\label{sec:Reentrant localization transition}
%%%%%%%%%%%%%%%%%%%%%%%%%%%%%%%%%%%%%%%%%%%%%%%%%%%%%%%%%%%%%%%%%%%%%%%%%%%%%%%%%%%%%%%%
%%%%%%%%%%%%%%%%%%%%%%%%%%%%%%%%%%%%%%%%%%%%%%%%%%%%%%%%%%%%%%%%%%%%%%%%%%%%%%%%%%%%%%%%
%%%%%%%%%%%%%%%%%%%%%%%%%%%%%%%%%%%%%%%%%%%%%%%%%%%%%%%%%%%%%%%%%%%%%%%%%%%%%%%%%%%%%%%%

%%%%%%%%%%%%%%%%%%%%%%%%%%%%%%%%%%%%%%%%%%%%%%%%%%%%%%%%%%%%%%%%%%%%%%%%%%%%%%%%%%%%%%%%
\subsection{Characterization}
%%%%%%%%%%%%%%%%%%%%%%%%%%%%%%%%%%%%%%%%%%%%%%%%%%%%%%%%%%%%%%%%%%%%%%%%%%%%%%%%%%%%%%%%

To characterize the localization properties of the dimerized quasiperiodic dipolar chain, we numerically diagonalize the Hamiltonian \eqref{eq:H} and compute various quantities of interest from its eigenvectors.

One of these quantities is the inverse participation ratio (IPR), defined as \cite{Bell1970,Thouless1974}
\begin{equation}
    \mathrm{IPR}(n) = \sum_{i=1}^{2\mathcal{N}}|\Psi_i(n)|^4
\label{eq:IPR}
\end{equation}
for an eigenstate $n$ with a $2\mathcal{N}$-component eigenvector $\mathbf{\Psi}(n)=\left(\Psi_1(n), \Psi_2(n), \ldots, \Psi_{2\mathcal{N}}(n)\right)$.
The IPR measures the localization degree of an eigenstate: It is equal to $1$ for a fully localized state, and scales as $\mathcal{N}^{-1}$ for an extended state.

Another measure of localization is the normalized participation ratio (NPR)
\begin{equation}
    \mathrm{NPR}(n) = \frac{1}{2\mathcal{N}}   \left(  \sum_{i=1}^{2\mathcal{N}}|\Psi_i(n)|^4 \right)^{-1}.
\label{eq:NPR}
\end{equation}
The NPR scales as $\mathcal{N}^{-1}$ for a localized state, while in a 1d chain with open boundary conditions as considered here it tends to $2/3$ for a fully extended state \cite{Thouless1974}.

Moreover, to help us identifying the presence of intermediate phases in the system, we use the quantity \cite{DasSarma2020}
\begin{equation}
    \eta = \log_{10}\left(  \langle \mathrm{IPR} \rangle \times \langle \mathrm{NPR} \rangle  \right).
\label{eq:eta}
\end{equation}
Here the notation $\langle \cdot \rangle$ denotes an averaging over a given fraction of the eigenspectrum.
If not stated otherwise, in this work we average over the complete set of eigenstates.
Since in, respectively, localized and extended phases, the averaged NPR and averaged IPR scale as $\mathcal{N}^{-1}$, the quantity $\eta < -\log_{10}\mathcal{N}$ in these phases.
In intermediate phases, however, as both the averaged NPR and averaged IPR remain finite for large system sizes, $\eta > -\log_{10}\mathcal{N}$. 
While we focus here on the above eigenvector-related quantities to monitor the localization properties of the system, we note that we also verified (not shown) that similar conclusions can be drawn from the distribution of the eigenvalues \cite{Oganesyan2007}.

%%%%%%%%%%%%%%%%%%%%%%%%%%%%%%%%%%%%%%%%%%%%%%%%%%%%%%%%%%%%%%%%%%%%%%%%%%%%%%%%%%%%%%%%
\subsection{Reentrant localization transition}
%%%%%%%%%%%%%%%%%%%%%%%%%%%%%%%%%%%%%%%%%%%%%%%%%%%%%%%%%%%%%%%%%%%%%%%%%%%%%%%%%%%%%%%%

\begin{figure*}[tb]
 \includegraphics[width=\linewidth]{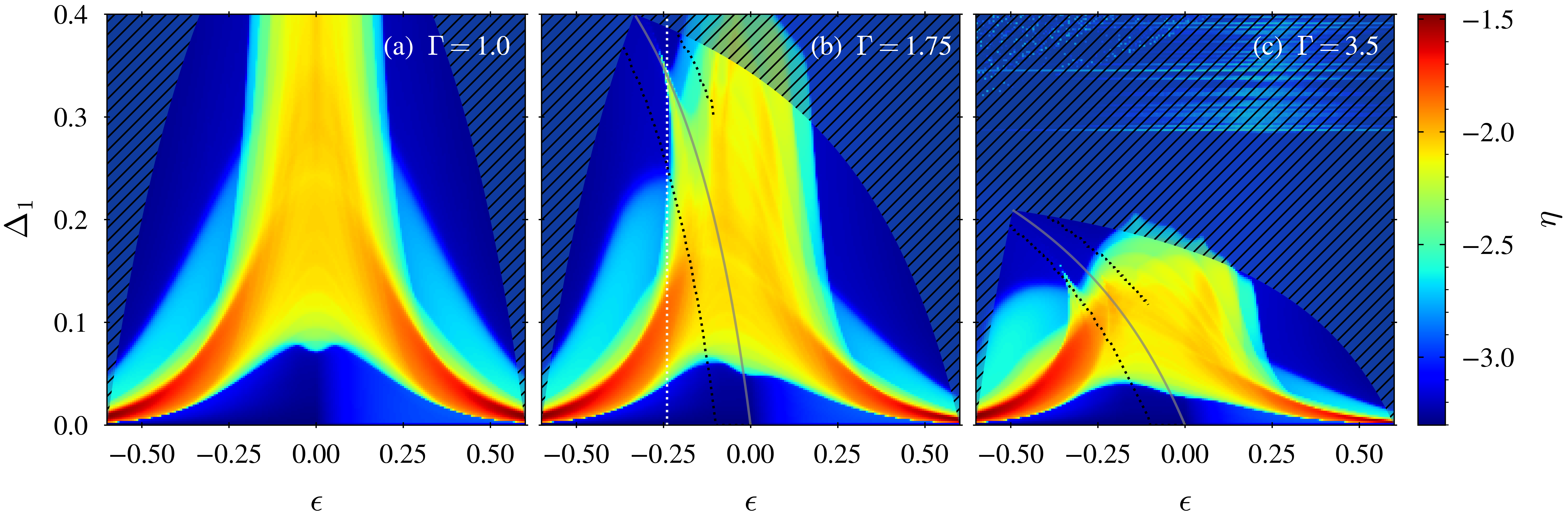}
 \caption{Localization phase diagrams of the model \eqref{eq:H}, showing the quantity $\eta$ [see Eq.~\eqref{eq:eta}] in the $(\epsilon,\Delta_1)$ plane.
 Increasing quasiperiodic strength ratios $\Gamma$ are considered from panel (a) to (c).
 The hatched areas represent parameter regions where the interemitter spacings $d_{1m}, d_{2m} \lesssim 3a$, which breaks down our dipolar approximation.
 A white dotted line in panel (b) highlights an RLT around $\epsilon=-0.24$.
 In panels (b) and (c), the solid grey lines correspond to the analytical estimate \eqref{eq:analytical curve}, while the surrounding black dotted lines are defined in the text.
 In the figure, the number of dimers $\mathcal{N}=1000$, and, as in the sequel of the paper, the incommensurate period $\beta = (\sqrt{13}+3)/2 + (\sqrt{5} + 1)/2$ and $d_1 + d_2 = 15a$.}
 \label{fig:phase diagrams eta}
\end{figure*}

To start our investigation of the localization properties of the Hamiltonian \eqref{eq:H}, we compute the quantity $\eta$ from Eq.~\eqref{eq:eta} as it allows us to easily monitor intermediate phases.
To find an RLT, we look for parameter regions, i.e., specific values of $\epsilon$ and $\Gamma$, for which several transitions from localized to intermediate phase appear as we increase the quasiperiodic modulation strengths.

We present in Fig.~\ref{fig:phase diagrams eta} our study of $\eta$ as a function of both $\epsilon$ and $\Delta_1$ for increasing values of $\Gamma$.
Small values of $\eta$, visible in blue, denote extended or localized phases, while larger values of $\eta$, visible from light blue to red, denote intermediate phases, hosting eigenstates of mixed natures. 
The black hatched areas in Fig.~\ref{fig:phase diagrams eta} represent regions of the parameter space that are forbidden by the constraint \eqref{eq:dipolar constraint}.

Figure \ref{fig:phase diagrams eta}(a) shows our results for $\eta$ for the case $\Gamma=1.0$ ($\Delta_2 = \Delta_1)$, i.e., for symmetric quasiperiodic modulations of the interemitter distances.
When the dimerization $\epsilon=0$, we recover results similar to the ones of the off-diagonal quasiperiodic AA model, that is the presence of a large intermediate phase. We note that in contrast to the usual off-diagonal AA model, here the all-to-all dipolar coupling makes this intermediate phase large but finite, as unveiled in Ref.~\cite{Hu2024}.
Dimerizing the chain with $|\epsilon| \neq 0$ produces the appearance of a transition to a localized phase, with an intermediate phase becoming thinner as $\epsilon$ increases.
The dimerization of the system thus competes against the intermediate phase.

We increase the quasiperiodic strength ratio to $\Gamma = 1.75$ ($\Delta_2 > \Delta_1$) in Fig.~\ref{fig:phase diagrams eta}(b).
Such an imbalance in the quasiperiodic modulations induces an asymmetry between the $\epsilon>0$ $(d_1>d_2)$ and $\epsilon<0$ $(d_1<d_2)$ cases.
In the latter case, where the interdimer distances $d_{2,m}$ are more modulated and on average larger than the intradimer ones $d_{1,m}$, we enter in the situation discussed in Sec.~\ref{sec:asymmetric modulation}, where a large enough modulation strength $\Delta_1$ allows a decrease of the local dimerization along the chain (see Fig.~\ref{fig:local epsilon}).
Interestingly, we observe in this regime a cusp-like shape of the intermediate phase around $\epsilon=-0.24$ (highlighted by a thin dotted line) and $\Delta_1 \sim 0.35$.
This corresponds to a clear signature of an RLT.
Indeed, increasing the quasiperiodic modulation for these particular values of $\epsilon$ leads the system to go from an extended to an intermediate and to a localized phase, and then to redelocalize itself into a second intermediate phase (within the cusp-like shape), to finally relocalize itself into a second localized phase.

To support the connection between this reentrant transition and a competition between the dimerization $\epsilon$ that tends to suppress the intermediate phase and a local undimerization induced by an increase of the asymmetric modulation, we plot in Fig.~\ref{fig:phase diagrams eta} as a grey line the values of $(\epsilon<0,\Delta_1)$ corresponding to $\textrm{min}(|\epsilon_m|)\simeq0$ [a configuration represented in Fig.~\ref{fig:spacing distributions}(c) as well as by the green lines in Fig.~\ref{fig:local epsilon}(a)].
These values are found analytically as
\begin{equation}
    \Delta_1^{\mathrm{c}}(\epsilon,\Gamma) = \frac{2\epsilon}{\epsilon(\Gamma+1)-(\Gamma-1)}.
\label{eq:analytical curve}
\end{equation}
To surround this simple analytical curve by an approximate region where the local undimerization effect is present, we also compute the fraction of dimers featuring $|\epsilon_m|\leqslant0.1$.
For a fixed $\epsilon$, the lower black dotted line corresponds to the modulation strength $\Delta_1$ for which at least one dimer has $|\epsilon_m|\leqslant0.1$, while the upper black dotted line is the modulation strength maximizing the fraction
of dimers with $|\epsilon_m|\leqslant0.1$.
As visible in Fig.~\ref{fig:phase diagrams eta}(b), although these lower and upper bounds are only approximate, they qualitatively predict the parameter region in which an RLT occurs in our system.
The local undimerization induced by the modulation thus provides an intuitive physical insight for why as well as for which parameters an RLT occurs in our system.

Before moving to a precise study of this RLT, we display in Fig.~\ref{fig:phase diagrams eta}(c) the case of $\Gamma = 3.5$.
By further increasing $\Gamma$, we increase the modulation on the distances $d_{2,m}$ as compared to the one on $d_{1,m}$, so that the parameter region respecting the constraint \eqref{eq:dipolar constraint} is reduced.
A similar cusp---and therefore an RLT---to that visible in Fig.~\ref{fig:phase diagrams eta}(b) is still present in Fig.~\ref{fig:phase diagrams eta}(c) for $\epsilon\sim-0.30$, now for a reduced value of $\Delta_1$.
As in Fig.~\ref{fig:phase diagrams eta}(b), the RLT occurs around values of $\epsilon$ and $\Delta_1$ where the local dimerization is reduced, a parameter region comprised between the lower and upper black dotted lines.
We note that considering values of the quasiperiodic strength ratio $\Gamma<1$ would simply reverse the roles of $\epsilon>0$ and $\epsilon<0$.
Indeed, the Hamiltonian \eqref{eq:H} is invariant under the transformation $(\Gamma, \epsilon) \rightarrow (1/\Gamma, -\epsilon)$.
Finally, other choices of $\beta$ may lead to qualitatively similar results, although it strongly modifies the parameter region in which an RLT appears, as unveiled in Ref.~\cite{Nair2025}.
In our study, while the sum of the bronze and golden ratio allows the observed RLT, the golden ratio leads only to a transition in an extremely narrow parameter region, supporting our hypothesis of an RLT originating from the local undimerization effect discussed previously.

\begin{figure}[tb]
 \includegraphics[width=.55\linewidth]{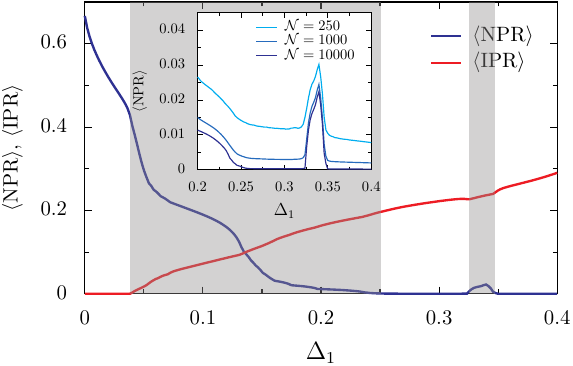}
 \caption{NPR and IPR, each averaged over the whole eigenspectrum, as a function of the quasiperiodic strength $\Delta_1$ for the choice of parameters indicated by the white dotted line in Fig.~\ref{fig:phase diagrams eta}(b), i.e., $\epsilon=-0.24$ and $\Gamma=1.75$.
 Parameter regions in which neither the NPR nor the IPR are zero, which indicate the presence of an intermediate phase, are highlighted in gray.
 The inset zooms in the second of such regions, which characterizes the RLT, and shows the averaged NPR for increasing system sizes.
 In the main figure, the number of dimer $\mathcal{N}=10000$.}
 \label{fig:NPR and IPR}
\end{figure}

We now move to the study of the RLT visible in Fig.~\ref{fig:phase diagrams eta}(b), and fix in the following $\Gamma=1.75$ and $\epsilon=-0.24$ (highlighted by the white dotted line in the figure).
We show in Fig.~\ref{fig:NPR and IPR} the averaged IPR and NPR [see Eqs.~\eqref{eq:IPR} and \eqref{eq:NPR}, respectively] for the latter values of parameters as a function of the quasiperiodic modulation strength $\Delta_1$.
Five distinguishable phases are clearly visible in the figure.
First, an averaged IPR of zero indicates an extended phase for $\Delta_1 \lesssim 0.04$.
As demonstrated by the fact that both the averaged NPR and IPR are nonzero, the system then enters into a first intermediate phase which lasts as long as $0.04 \lesssim \Delta_1 \lesssim 0.25$.
It is followed by a first localized phase, indicated by $\langle \mathrm{NPR} \rangle = 0$ for $0.25 \lesssim \Delta_1 \lesssim 0.32$.
A further increase of the quasiperiodic modulation leads to the entrance into a second intermediate phase characterizing the RLT, for $0.32 \lesssim \Delta_1 \lesssim 0.35$.
It is then followed by a second localized phase, for $\Delta_1 \gtrsim 0.35$.
We zoom in around the second intermediate phase in the inset of Fig.~\ref{fig:NPR and IPR}, in which the averaged NPR is shown for increasing system sizes.
A significant peak in $\langle \mathrm{NPR} \rangle$ is visible for $\mathcal{N}=250$, demonstrating an RLT already for small system sizes.

\begin{figure}[t]
 \includegraphics[width=.55\linewidth]{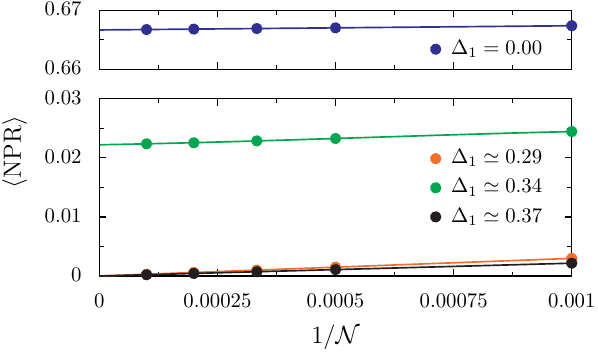}
 \caption{NPR [see Eq.~\eqref{eq:NPR}] averaged over the whole eigenspectrum as a function of the inverse number of dimers $1/\mathcal{N}$. Increasing values of the quasiperiodic strength $\Delta_1$ are shown, demonstrating the finite value of the NPR as the system size tends to infinity when $\Delta_1\simeq0.34$.
 Other parameters are the same as in Fig.~\ref{fig:NPR and IPR}.}
 \label{fig:NPR scaling}
\end{figure}

To ensure that the aforementioned reentrant transition is not a finite size effect, we study the scaling of the averaged NPR with the system size $\mathcal{N}$ in Fig.~\ref{fig:NPR scaling}.
Several values of the quasiperiodic modulation strength $\Delta_1$ are considered, showing the different phases of the system.
In the perfectly extended phase where the quasiperiodic strength $\Delta_1=0.00$ (blue), the averaged NPR is constant and equal to $2/3$ as expected.
In the first localized phase (orange), as well as in the second one (black), $\langle \mathrm{NPR} \rangle \to 0$ as $\mathcal{N} \to \infty$, a scaling that is an unambiguous marker of the localized nature of the system eigenstates.
Moreover it is clear from Fig.~\ref{fig:NPR scaling} that within the second intermediate phase (green), the averaged NPR tends to a finite value when $\mathcal{N} \to \infty$, ensuring the fact that it is a genuine intermediate phase.

\begin{figure}[tb]
 \includegraphics[width=.5\linewidth]{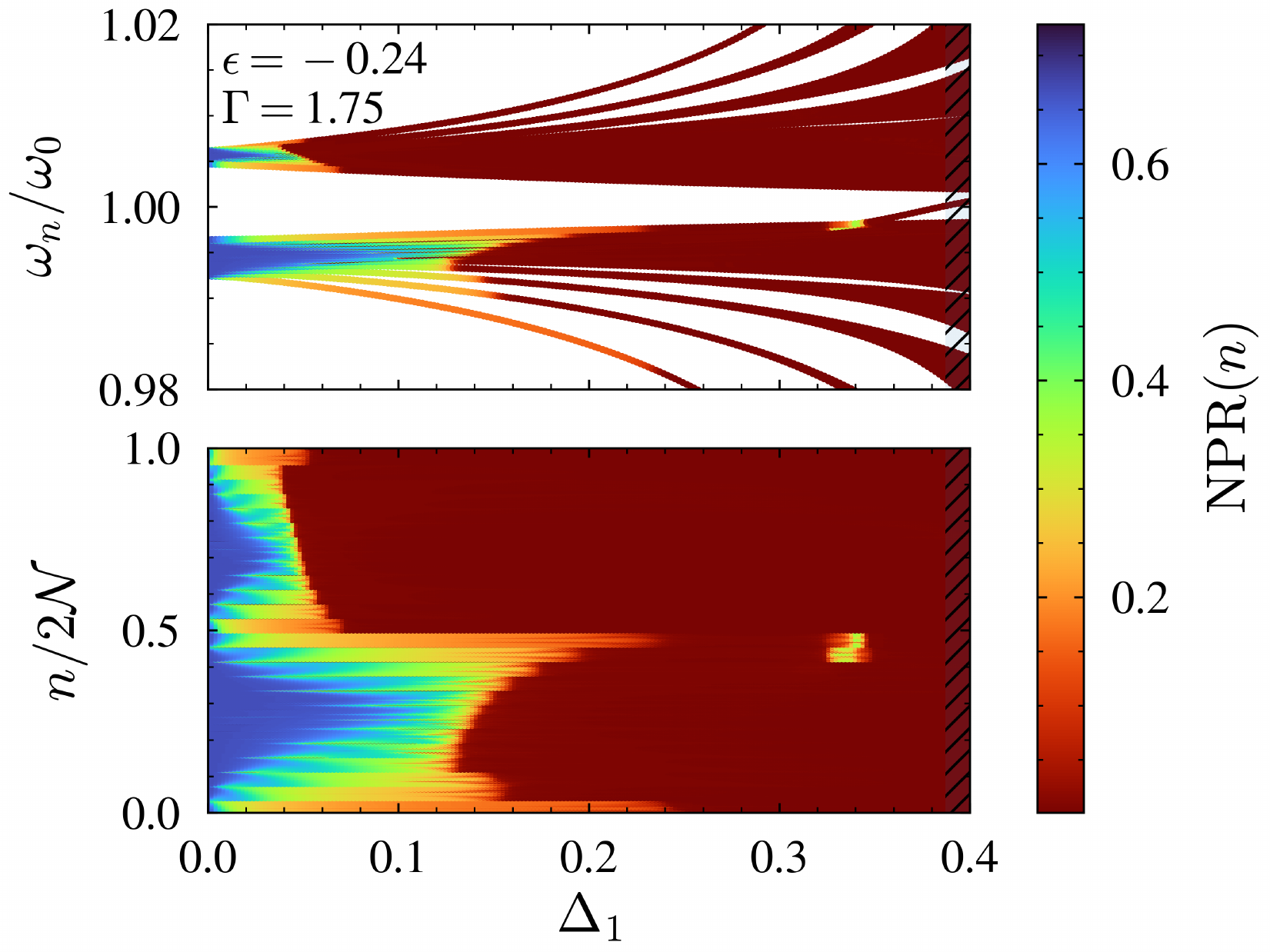}
 \caption{Eigenspectrum as a function of the quasiperiodic strength $\Delta_1$.
 The color code indicates the NPR [see Eq.~\eqref{eq:NPR}] of each eigenstate $n$.
 Upper panel: Eigenfrequencies $\omega_n$ in units of the individual emitter frequency $\omega_0$.
 Lower panel: Sorted normalized index $n/2\mathcal{N}$.
 An RLT can be observed for the eigenstates near the middle of the spectrum around $\Delta_1 \simeq 0.34$.
As in Fig.~\ref{fig:phase diagrams eta}, the hatched area corresponds to the parameter region contravening the condition \eqref{eq:dipolar constraint}.
 In the figure, the number of dimers $\mathcal{N}=1000$, and other parameters are the same as in Figs.~\ref{fig:NPR and IPR} and \ref{fig:NPR scaling}.}
 \label{fig:spectrum vs Delta1}
\end{figure}

So far, we have only looked at quantities averaged over the entire eigenspectrum.
To better understand how the Hamiltonian \eqref{eq:H} is affected by the RLT, we show in Fig.~\ref{fig:spectrum vs Delta1} its corresponding  eigenspectrum as a function of the quasiperiodic modulation $\Delta_1$, and present in a color code the NPR for each eigenstate $n$ of the system.
A red (blue) color indicates localized (extended) states, while other colors denote states with an intermediate value of the participation ratio.
In the upper panel, we show the ordered eigenfrequencies in units of the bare emitter frequency, $\omega_n/\omega_0$, while in the lower panel we represent the normalized eigenstate index $n/2\mathcal{N}$.

A clear asymmetry is visible between the two energy bands of the spectrum.
It originates from the all-to-all dipolar coupling between the emitters which breaks the chiral symmetry in the Hamiltonian \eqref{eq:H}.
For the longitudinally polarized dipoles we consider in this work, the all-to-all coupling induces a low-energy band with a larger bandwidth than that of the high-frequency one \cite{Downing2018}.
When increasing the modulation $\Delta_1$, we observe in Fig.~\ref{fig:spectrum vs Delta1} the characteristic formation of minibands in quasiperiodic systems.
Moreover, from the bandwidth asymmetry between the two energy bands, the high-energy one enters in a localized phase sooner than the low-energy one.
Interestingly, the RLT is observed only for eigenstates belonging to the upper edge of the low-energy band ($0.4 \lesssim n/2\mathcal{N} \lesssim 0.5$), and arises once such upper edge breaks into the bandgap.
As visible in the lower panel of Fig.~\ref{fig:spectrum vs Delta1}, around $\unit[10]{\%}$ of the eigenstates of the system undergo an RLT.
An additional study including finite size scaling and multifractal analysis presented in Appendix \ref{sec:app:Multifractal analysis} demonstrates the critical nature of this portion of the eigenspectrum in the second intermediate phase.

We note that when considering only nearest-neighbor couplings, as discussed in Appendix~\ref{sec:app:Nearest-neighbor}, the bandstructure recovers its mirror symmetry around $\omega_0$.
This implies that in this case the RLT is observed for eigenstates belonging to both the upper and lower bands.
The all-to-all dipolar coupling, by breaking the chiral symmetry of the Hamiltonian, halves the fraction of the eigenspectrum entering in a second intermediate phase and is hence clearly detrimental to the RLT, in agreement with the results of Ref.~\cite{Wang2023} on next-to-nearest neighbor couplings.

For the longitudinally polarized dipoles which we consider in this work, the low-energy band is bright in the sense that it couples to the vacuum electromagnetic field, while the high-frequency one is dark \cite{Downing2018}.
Moreover, we verified that the very same physics appears when considering transversally polarized dipoles, except that the eigenstates undergoing the RLT now belong to the high-energy band, which, for transverse dipoles, is again the bright one.

\begin{figure}[b]
 \includegraphics[width=.5\linewidth]{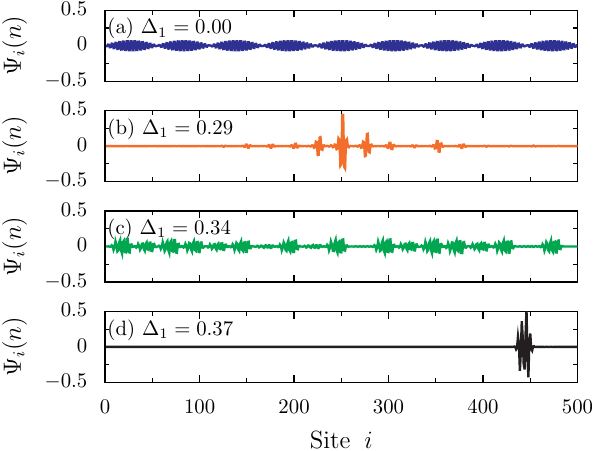}
 \caption{Eigenvectors $\Psi_i(n)$ along the sites $i$ of a chain composed of $\mathcal{N}=250$ dimers, for increasing quasiperiodic modulation strengths.
 In each panel, the eigenstate index of the plotted state is $n=241$.
 Other parameters are the same as in Figs.~\ref{fig:NPR and IPR} to \ref{fig:spectrum vs Delta1}.}
 \label{fig:eigenvectors}
\end{figure}

To conclude our analysis of the localization properties of dimerized quasiperiodic dipolar chains, we examine the precise shape of the eigenvectors undergoing the RLT.
To this end, we compute, along the sites $i$ of a chain of $\mathcal{N}=250$ dimers, the eigenvector $\Psi_i(n)$ corresponding to the state $n=241$ (i.e., $n/2\mathcal{N}=0.482$).
We present the results in Fig.~\ref{fig:eigenvectors} for increasing modulation strength from panel (a) to (d).
While in the perfectly extended phase the eigenstate has a sinusoidal envelope (blue line), it becomes exponentially localized around a few sites in the first localized phase (orange line).
However, due to the RLT, increasing the quasiperiodic modulation $\Delta_1$ leads the system to enter in a second intermediate phase.
Here a delocalization is visible, as the eigenstate is now spread over a large number of sites (green line).
Eventually, increasing further $\Delta_1$, the state becomes exponentially localized on a few sites again (black line).
In the next section, we  examine how this counterintuitive eigenstate delocalization induced by an increase of the quasiperiodic modulation could be probed through transport experiments, taking into account the inherent system losses.

%%%%%%%%%%%%%%%%%%%%%%%%%%%%%%%%%%%%%%%%%%%%%%%%%%%%%%%%%%%%%%%%%%%%%%%%%%%%%%%%%%%%%%%%
%%%%%%%%%%%%%%%%%%%%%%%%%%%%%%%%%%%%%%%%%%%%%%%%%%%%%%%%%%%%%%%%%%%%%%%%%%%%%%%%%%%%%%%%
%%%%%%%%%%%%%%%%%%%%%%%%%%%%%%%%%%%%%%%%%%%%%%%%%%%%%%%%%%%%%%%%%%%%%%%%%%%%%%%%%%%%%%%%
\section{Transport simulation with lossy emitters}
\label{sec:Transport}
%%%%%%%%%%%%%%%%%%%%%%%%%%%%%%%%%%%%%%%%%%%%%%%%%%%%%%%%%%%%%%%%%%%%%%%%%%%%%%%%%%%%%%%%
%%%%%%%%%%%%%%%%%%%%%%%%%%%%%%%%%%%%%%%%%%%%%%%%%%%%%%%%%%%%%%%%%%%%%%%%%%%%%%%%%%%%%%%%
%%%%%%%%%%%%%%%%%%%%%%%%%%%%%%%%%%%%%%%%%%%%%%%%%%%%%%%%%%%%%%%%%%%%%%%%%%%%%%%%%%%%%%%%

In Sec.~\ref{sec:Reentrant localization transition} we unveiled an anomalous transition from localized to critical eigenstates when increasing the quasiperiodic modulation strength.
Interestingly, from a transport perspective such a reentrant transition could imply quasiperiodic modulation-enhanced transport, an intriguing mechanism that has recently been extensively studied for random disorder in the field of strongly coupled light-matter systems \cite{Chavez21,Allard2022}.
To assess whether the reentrant transition can imply an enhancement of the propagation as the quasiperiodic modulation strength is increased, and whether this could be probed taking into account the inherent losses of dipolar emitters, we simulate the transport properties of the dimerized quasiperiodic dipolar chain in a driven-dissipative scenario.

For that purpose, we add to the Hamiltonian \eqref{eq:H} the driving term
\begin{equation}
    H_\mathrm{drive}(t) = \hbar\Omega_\mathrm{R}\sin(\omega_\mathrm{d}t)\left( a^{\phantom{\dagger}}_1 + a^{\dagger}_1 \right)
    \label{eq: Hdrive}
\end{equation}
modeling an electric field continuously acting on the first emitter (i.e., in the dimer $m=1$ and sublattice $A$) at a driving frequency $\omega_\mathrm{d}$.
Here, $\Omega_\mathrm{R} = E_0 \sqrt{Q^2/2M\hbar\omega_0}$ is the Rabi frequency where $E_0$ is the field amplitude.
We then assume that the transport dynamics is described by the Lindblad master equation for the density matrix $\rho$, i.e.,
\begin{equation}
    \dot{\rho} = \frac{\mathrm{i}}{\hbar}\left[ \rho, H + H_{\mathrm{drive}}(t) \right] 
    - \frac{\gamma}{2}\sum_{m=1}^{\mathcal{N}} \left(  \left\{ a_m^\dagger a_m^{\phantom{\dagger}} + b_m^\dagger b_m^{\phantom{\dagger}}, \rho \right\}   - 2a_m^{\phantom{\dagger}} \rho a_m^\dagger - 2b_m^{\phantom{\dagger}} \rho b_m^\dagger \right).
    \label{eq:Lindblad master equation}
\end{equation}
In this open quantum system approach, the damping rate $\gamma$ quantifies the dissipation of the dipolar emitters into a phenomenological Markovian bath.
Typical dissipation mechanisms are Ohmic losses or radiative damping.

To study the transport properties along the chain of dipoles, we introduce the dimensionless dipole moment $p^A_m = \langle a_m + a_m^\dagger \rangle_\rho$ ($p^B_m = \langle b_m + b_m^\dagger \rangle_\rho$) bared by a dipole belonging to the dimer $m$ and sublattice $A$ ($B$).
Here the notation $\langle \mathcal{O} \rangle_\rho=\mathrm{Tr}(\rho\mathcal{O})$ denotes the trace of the operator $\mathcal{O}$ over the density operator.
We note that such dimensionless dipole moments are related to the power radiated by a dipole in the far field through the classical Larmor formula \cite{Jackson2007}.

\begin{figure}[tb]
 \includegraphics[width=.5\linewidth]{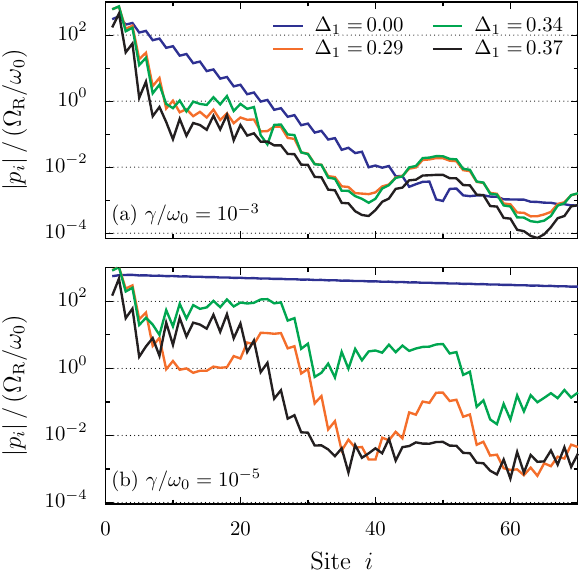}
 \caption{Steady-state amplitude of the dipole moment $|p_i|$ on site $i$ in units of the dimensionless Rabi frequency $\Omega_\mathrm{R}/\omega_0$, along the first $70$ sites of a chain composed of $500$ emitters.
 Results corresponding to increasing quasiperiodic strengths are shown, and damping rates $\gamma/\omega_0=10^{-3}$ and $\gamma/\omega_0=10^{-5}$ are considered in panels (a) and (b), respectively.
 The driving frequencies are chosen to correspond to the eigenfrequencies of the eigenstates shown in Fig.~\ref{fig:eigenvectors}.
 Other parameters are the same as in Figs.~\ref{fig:NPR and IPR} to \ref{fig:eigenvectors}.
 }
 \label{fig:transport}
\end{figure}

We numerically compute the steady-state amplitudes of the dipole moments $p_m^A$ and $p_m^B$, and recast them into the site-dependent quantity $|p_i|$.
The result is presented as log-linear plots in Fig.~\ref{fig:transport} for the first $70$ sites of a chain of $250$ dimers.
To compare such driven-dissipative simulations to our previous lossless results, we choose a driving frequency that corresponds precisely to the eigenfrequency of the states shown in Fig.~\ref{fig:eigenvectors}, and consider the same increasing modulation strengths.

Figure \ref{fig:transport}(a) displays the case of a damping rate $\gamma/\omega_0=10^{-3}$, a value that could be achieved experimentally using low-loss emitters such as, e.g., microwave antennas or SiC nanoparticles \cite{Mann2018,Wang2018b}.
First looking at the periodic case ($\Delta_1=0.00$), we observe that the propagation along the chain consists in two regimes, namely, an exponential decay (visible as a straight line in such a log-linear plot), followed by another regime starting around the site $i=50$. We verified that this second regime corresponds in fact to an algebraic decay that originates directly from the all-to-all dipolar coupling \eqref{eq:Omega}.
Once the interdipole distances are modulated quasiperiodically ($\Delta_1 \neq 0$), the propagation along the first sites of the chain is drastically reduced, with an exponential decay that becomes steeper.
The algebraic decay, on the other hand, is replaced by decaying oscillations.
As the modulation strength is increased to go from driving a localized (orange line) to a critical (green line) state, it is observed that this scenario of damping rate does not allow the RLT to be easily detectable.
Indeed, the propagation corresponding to the two states are very similar along most of the chain.
Nevertheless, a slight threefold enhancement of the dipole moment amplitude is visible around the sites $10$ to $20$.
We note that choosing wisely the excitation site of the driving \eqref{eq: Hdrive} by searching where the critical eigenstate in Fig.~\ref{fig:eigenvectors}(c) is most localized may allow a better probe of the RLT. 

We next consider the scenario of a narrower linewidth in Fig.~\ref{fig:transport}(b), with $\gamma/\omega_0=10^{-5}$.
Here we observe a large enhancement of the transport characteristics of the system as we increase the quasiperiodic modulation strength from $\Delta_1=0.29$ [orange line, corresponding to the localized state of Fig.~\ref{fig:eigenvectors}(b)] to $\Delta_1=0.34$ [green line, corresponding to the critical state of Fig.~\ref{fig:eigenvectors}(c)], with up to a hundredfold increase in the dipole amplitudes, especially at long distances.
Increasing further the modulation to $\Delta_1=0.37$ [black line, corresponding to the localized state of Fig.~\ref{fig:eigenvectors}(d)], an overall reduction of the propagation is visible, again especially at long distances. 
However, the necessity of such a small damping rate to clearly observe these effects reveals the limited robustness of the RLT to losses.
We note that similar conclusions have been drawn when adding non-Hermiticity to an AA model with staggered potentials \cite{Jiang2021}.

%%%%%%%%%%%%%%%%%%%%%%%%%%%%%%%%%%%%%%%%%%%%%%%%%%%%%%%%%%%%%%%%%%%%%%%%%%%%%%%%%%%%%%%%
%%%%%%%%%%%%%%%%%%%%%%%%%%%%%%%%%%%%%%%%%%%%%%%%%%%%%%%%%%%%%%%%%%%%%%%%%%%%%%%%%%%%%%%%
%%%%%%%%%%%%%%%%%%%%%%%%%%%%%%%%%%%%%%%%%%%%%%%%%%%%%%%%%%%%%%%%%%%%%%%%%%%%%%%%%%%%%%%%
\section{Conclusion}
\label{sec:Conclusion}
%%%%%%%%%%%%%%%%%%%%%%%%%%%%%%%%%%%%%%%%%%%%%%%%%%%%%%%%%%%%%%%%%%%%%%%%%%%%%%%%%%%%%%%%
%%%%%%%%%%%%%%%%%%%%%%%%%%%%%%%%%%%%%%%%%%%%%%%%%%%%%%%%%%%%%%%%%%%%%%%%%%%%%%%%%%%%%%%%
%%%%%%%%%%%%%%%%%%%%%%%%%%%%%%%%%%%%%%%%%%%%%%%%%%%%%%%%%%%%%%%%%%%%%%%%%%%%%%%%%%%%%%%%

To summarize, we explored the localization properties of a dimerized chain of dipolar emitters whose interemitter distances are modulated quasiperiodically.
In particular, we investigated the fate of RLTs, anomalous transitions in which eigenstates undergo a transition from localized to critical as the strength of quasiperiodic modulation is increased.
While RLTs have been predicted in several quasiperiodic systems \cite{Ramakumar2015,Goblot2020, Roy2021, Jiang2021, Wu2021, Zuo2022, Han2022, Padhan2022, Wang2023, Qi2023,Roy2022, Guan2023, Dai2023, Goncalves2023, Goncalves2023, Vaidya2023, Shimasaki2024, Miranda2024, Xu2024, Guo2024, Padhan2024, Ganguly2024, Tabanelli2024, Li_PRR2024, Lu2025, Li2023_arxiv, Li2023_arxiv2, Nair2025, Chang2025}, their origin as well as their extent and robustness to system's complexities remain elusive.
Here, we unveiled that in a realistic dipolar system, the interplay between a chain dimerization and an asymmetric quasiperiodic modulation of the dipole spacings induces the revival of an intermediate phase, leading to an RLT.
We hence demonstrate the robustness of this phenomenon to all-to-all Coulomb interactions, despite their detrimental effect on the transition.

Furthermore, we studied the impact of dissipation on RLTs using an open quantum system approach to conduct transport simulations.
This allowed us to demonstrate the manifestation of quasiperiodic modulation-enhanced transport in the context of low-loss emitters.
Importantly, these simulations also revealed the detrimental impact of dissipation on the RLT, illustrating the fragile nature of this anomalous transition.

Along with recent works \cite{Wang2021, Hu2024, Citrin2024}, our study represents a further step towards understanding the localization properties of dipolar quasiperiodic systems, platforms that are of particular interest to investigate the impact of long-range interactions as well as non-Hermiticity.
Numerous avenues remain open in this developing field.
Notably, the possible interplay between the RLT and its coupling to electromagnetic modes
could lead to a rich phenomenology \cite{Macedo2025}.
Moreover, the exotic topological properties of disordered dimerized chains such as standard topological Anderson insulator phases \cite{Meier2018,Longhi2020} as well as ungapped ones \cite{Ren2024} could have unexpected features in quasiperiodic dipolar systems.

%%%%%%%%%%%%%%%%%%%%%%%%%%%%%%%%%%%%%%%%%%%%%%%%%%%%%%%%%%%%%%%%%%%%%%%%%%%%%%%%%%%%%%%%
%%%%%%%%%%%%%%%%%%%%%%%%%%%%%%%%%%%%%%%%%%%%%%%%%%%%%%%%%%%%%%%%%%%%%%%%%%%%%%%%%%%%%%%%
%%%%%%%%%%%%%%%%%%%%%%%%%%%%%%%%%%%%%%%%%%%%%%%%%%%%%%%%%%%%%%%%%%%%%%%%%%%%%%%%%%%%%%%%
\begin{acknowledgments}
This work of the Interdisciplinary Thematic Institute QMat, as part of the ITI 2021-2028 program of the University of Strasbourg, CNRS, and Inserm, was supported by IdEx Unistra (ANR 10 IDEX 0002), and by SFRI STRAT’US Projects No.\ ANR-20-SFRI-0012 and No.\ ANR-17-EURE- 0024 under the framework of the French Investments for the Future Program.
\end{acknowledgments}
%%%%%%%%%%%%%%%%%%%%%%%%%%%%%%%%%%%%%%%%%%%%%%%%%%%%%%%%%%%%%%%%%%%%%%%%%%%%%%%%%%%%%%%%
%%%%%%%%%%%%%%%%%%%%%%%%%%%%%%%%%%%%%%%%%%%%%%%%%%%%%%%%%%%%%%%%%%%%%%%%%%%%%%%%%%%%%%%%
%%%%%%%%%%%%%%%%%%%%%%%%%%%%%%%%%%%%%%%%%%%%%%%%%%%%%%%%%%%%%%%%%%%%%%%%%%%%%%%%%%%%%%%%

\appendix

%%%%%%%%%%%%%%%%%%%%%%%%%%%%%%%%%%%%%%%%%%%%%%%%%%%%%%%%%%%%%%%%%%%%%%%%%%%%%%%%%%%%%%%%
%%%%%%%%%%%%%%%%%%%%%%%%%%%%%%%%%%%%%%%%%%%%%%%%%%%%%%%%%%%%%%%%%%%%%%%%%%%%%%%%%%%%%%%%
%%%%%%%%%%%%%%%%%%%%%%%%%%%%%%%%%%%%%%%%%%%%%%%%%%%%%%%%%%%%%%%%%%%%%%%%%%%%%%%%%%%%%%%%
\section{Second quantized Hamiltonian}
\label{sec:app:Second quantized Hamiltonian}
%%%%%%%%%%%%%%%%%%%%%%%%%%%%%%%%%%%%%%%%%%%%%%%%%%%%%%%%%%%%%%%%%%%%%%%%%%%%%%%%%%%%%%%%
%%%%%%%%%%%%%%%%%%%%%%%%%%%%%%%%%%%%%%%%%%%%%%%%%%%%%%%%%%%%%%%%%%%%%%%%%%%%%%%%%%%%%%%%
%%%%%%%%%%%%%%%%%%%%%%%%%%%%%%%%%%%%%%%%%%%%%%%%%%%%%%%%%%%%%%%%%%%%%%%%%%%%%%%%%%%%%%%%
In this Appendix, we provide a detailed derivation of the second quantized Hamiltonian $H$ of Eq.~\eqref{eq:H}. The dipolar system introduced in Fig.~\ref{fig:sketch} can be described by 
\begin{equation}
\label{eq:app:H}
H=\sum_{m=1}^{\mathcal{N}}\sum_{s=A,B}
\left[
\frac{{(\boldsymbol{\Pi}_m^s)}^2}{2M}+\frac{M\omega_0^2}{2}{{(\mathbf{h}_m^s)}^2}\right]
+\frac 12 \sum_{\substack{m,m',s,s'\\(m,s)\neq(m',s')}}\frac{\mathbf{p}_m^s\cdot\mathbf{p}_{m'}^{s'}-3(\mathbf{p}_m^s\cdot\hat z)(\mathbf{p}_{m'}^{s'}\cdot\hat z)}{{(r_{m,m'}^{s,s'})}^3}.
\end{equation}
In the above equation, the first term (in square brackets) corresponds to a set of $2\mathcal{N}$ uncoupled 
dipoles, with displacement field $\mathbf{h}_m^s$ and conjugate momentum $\boldsymbol{\Pi}_m^s$.
The second term represents the quasistatic dipole-dipole interaction, with electric dipole moment 
$\mathbf{p}_m^s=-Q\mathbf{h}_m^s$. The other quantities in Eq.~\eqref{eq:app:H} are defined in 
Sec.~\ref{sec:Model:Model} of the main text.

For longitudinally polarized dipoles along the $z$ direction, the Hamiltonian \eqref{eq:app:H} reduces to
\begin{equation}
\label{eq:app:H:H}
H=\sum_{m=1}^{\mathcal{N}}\sum_{s=A,B}
\left[
\frac{(\Pi_m^s)^2}{2M}+\frac{M\omega_0^2}{2}{({h}_m^s)^2}\right]
-Q^2\!\!\!\! \sum_{\substack{m,m',s,s'\\(m,s)\neq(m',s')}}\frac{{h}_m^s {h}_{m'}^{s'}}{(r_{m,m'}^{s,s'})^3}.
\end{equation}
Introducing the bosonic ladder operators
\begin{subequations}
\begin{align}
a_m&=\sqrt{\frac{M\omega_0}{2\hbar}}\, h_m^A+\frac{\mathrm{i}}{\sqrt{2\hbar M\omega_0}}\, \Pi_m^A, \\
b_m&=\sqrt{\frac{M\omega_0}{2\hbar}}\, h_m^B+\frac{\mathrm{i}}{\sqrt{2\hbar M\omega_0}}\, \Pi_m^B, 
\end{align}
\end{subequations} 
Eq.~\eqref{eq:app:H:H} yields the Hamiltonian \eqref{eq:H}.

%%%%%%%%%%%%%%%%%%%%%%%%%%%%%%%%%%%%%%%%%%%%%%%%%%%%%%%%%%%%%%%%%%%%%%%%%%%%%%%%%%%%%%%%
%%%%%%%%%%%%%%%%%%%%%%%%%%%%%%%%%%%%%%%%%%%%%%%%%%%%%%%%%%%%%%%%%%%%%%%%%%%%%%%%%%%%%%%%
%%%%%%%%%%%%%%%%%%%%%%%%%%%%%%%%%%%%%%%%%%%%%%%%%%%%%%%%%%%%%%%%%%%%%%%%%%%%%%%%%%%%%%%%
\section{Multifractal analysis}
\label{sec:app:Multifractal analysis}
%%%%%%%%%%%%%%%%%%%%%%%%%%%%%%%%%%%%%%%%%%%%%%%%%%%%%%%%%%%%%%%%%%%%%%%%%%%%%%%%%%%%%%%%
%%%%%%%%%%%%%%%%%%%%%%%%%%%%%%%%%%%%%%%%%%%%%%%%%%%%%%%%%%%%%%%%%%%%%%%%%%%%%%%%%%%%%%%%
%%%%%%%%%%%%%%%%%%%%%%%%%%%%%%%%%%%%%%%%%%%%%%%%%%%%%%%%%%%%%%%%%%%%%%%%%%%%%%%%%%%%%%%%

In the main text, we assess the presence of an intermediate phase through the computation of the IPR, NPR, as well as of the quantity $\eta$ [see, respectively, Eqs.~\eqref{eq:IPR}, \eqref{eq:NPR}, and \eqref{eq:eta}].
The nonzero value of both the averaged IPR and NPR, accompanied by a value of $\eta > -\log_{10}\mathcal{N}$ suggests the existence of such a phase.
A key property of RLTs is also that the eigenstates undergoing the transition become critical \cite{Roy2022}.
To distinguish these critical eigenstates from localized or extended states, a widely used method is to look at their multifractal characteristics \cite{Mirlin_RevModPhys}.

In this Appendix, we conduct a multifractal analysis to ensure the critical nature of the eigenstates in the reentrant intermediate phase which we observe in the main text (see the yellow spot around $\Delta_1 \simeq 0.34$ in Fig.~\ref{fig:spectrum vs Delta1}(b), for eigenstates with indices $n/2\mathcal{N} \in [0.4, 0.5]$).
To this end, we compute the $q$-dependent generalized IPR defined as \cite{Mirlin_RevModPhys}
\begin{equation}
    \mathrm{IPR}_q(n) = \sum_{i=1}^{2\mathcal{N}}|\Psi_i(n)|^{2q} \underset{\mathcal{N}\to\infty}{\sim} \mathcal{N}^{-\tau_{q}(n)}
    \label{eq:generalized IPR}
\end{equation}
and analyze its scaling with the system size to extract the multifractal exponent $\tau_q$, as done, e.g., in Ref.~\cite{Roy2022}.
Localized and extended states are characterized, respectively, by a multifractal exponent $\tau_q=0$ and $\tau_q=q-1$.
Any other behavior of the multifractal exponent as a function of $q$ indicates multifractality \cite{Mirlin_RevModPhys}.
We note that the case $q=2$ corresponds to the usual IPR, as defined in Eq.~\eqref{eq:IPR}.

\begin{figure}
 \includegraphics[width=\linewidth]{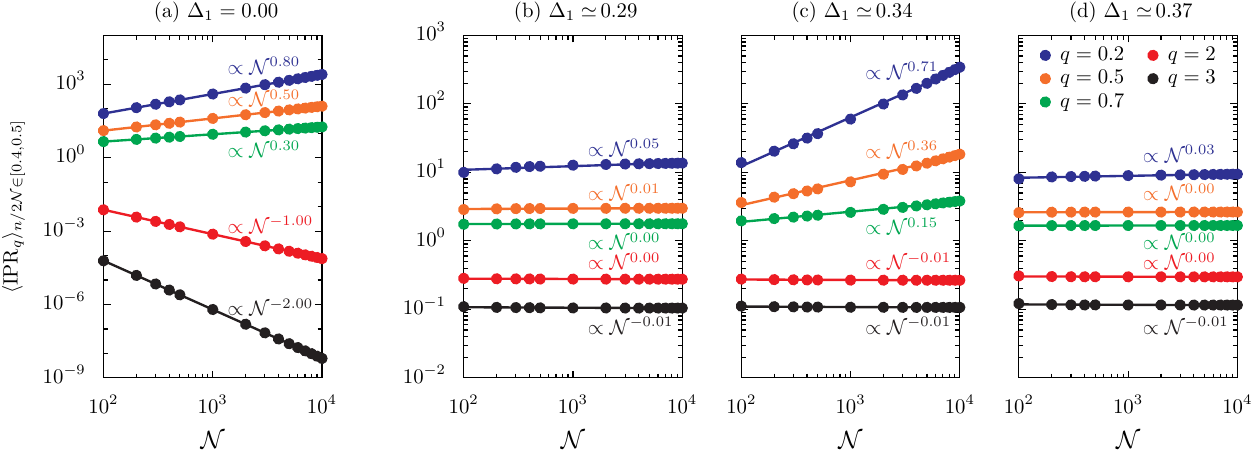}
 \caption{Scaling of the generalized IPR [see Eq.~\eqref{eq:generalized IPR}] with the system size $\mathcal{N}$.
 Increasing values of the quasiperiodic modulation strength are considered from panels (a) to (d).
 For each of them, the average of the generalized IPR over the eigenstates with indices $n/2\mathcal{N} \in [0.4,0.5]$ is plotted for different values of the exponent $q$, considering system sizes from $\mathcal{N}=100$ to $\mathcal{N}=10000$.
 A linear regression then allows us to find the scaling law $\mathcal{N}^{-\tau_q}$.
 In the figure, the dimerization $\epsilon=-0.24$ and the quasiperiodic strength ratio $\Gamma=1.75$, the same parameters as considered in most of the main text.}
 \label{fig:IPRq scaling}
\end{figure}

The result of a computation of the average of the generalized IPR \eqref{eq:generalized IPR} over the eigenstates concerned by the RLT, namely, the ones with indices $n/2\mathcal{N} \in [0.4,0.5]$ is presented in Fig.~\ref{fig:IPRq scaling} for increasing values of the quasiperiodic modulation strength $\Delta_1$ and for values of $q$ between $0.2$ and $3$.
A linear regression considering large system sizes from $\mathcal{N}=100$ to $\mathcal{N}=10000$ allows us to identify the value of $\tau_q$.

In Fig.~\ref{fig:IPRq scaling}(a) we present the case of a periodic chain.
As expected, since all the eigenstates are here extended, the generalized IPR scales with $\mathcal{N}^{-(q-1)}$.
Increasing the quasiperiodic modulation strength to $\Delta_1\simeq0.29$ in Fig.~\ref{fig:IPRq scaling}(b), we arrive at what we consider in the main text as localized eigenstates.
Our study of the generalized IPR confirms this characterization, as the latter stays constant when increasing the system size no matter the value of $q$, up to small finite size effects for small $q$.
The same conclusion can be drawn from Fig.~\ref{fig:IPRq scaling}(d), when a modulation strength $\Delta_1\simeq0.37$ is considered.
In between these two localized phases, however, we observe in Fig.~\ref{fig:IPRq scaling}(c) that the multifractal exponent has a nontrivial dependence on $q$ when $\Delta_1\simeq0.34$.
Indeed, for such a modulation strength, the generalized IPR averaged over the eigenstates concerned by the RLT increases with the system size for small values of $q$, but remains constant for larger ones.
This is consistent with what we observe in the main text and confirms that the second intermediate phase, present in the system for modulation strengths $0.32 \lesssim \Delta_1 \lesssim 0.35$, indeed host critical eigenstates.
Consequently, the transition between the latter phase and the second localized phase [visible here in Fig.~\ref{fig:IPRq scaling}(d)] is indeed a reentrant localization transition.

%%%%%%%%%%%%%%%%%%%%%%%%%%%%%%%%%%%%%%%%%%%%%%%%%%%%%%%%%%%%%%%%%%%%%%%%%%%%%%%%%%%%%%%%
%%%%%%%%%%%%%%%%%%%%%%%%%%%%%%%%%%%%%%%%%%%%%%%%%%%%%%%%%%%%%%%%%%%%%%%%%%%%%%%%%%%%%%%%
%%%%%%%%%%%%%%%%%%%%%%%%%%%%%%%%%%%%%%%%%%%%%%%%%%%%%%%%%%%%%%%%%%%%%%%%%%%%%%%%%%%%%%%%
\section{Nearest-neighbor chain}
\label{sec:app:Nearest-neighbor}
\begin{figure}[b]
 \includegraphics[width=.63\linewidth]{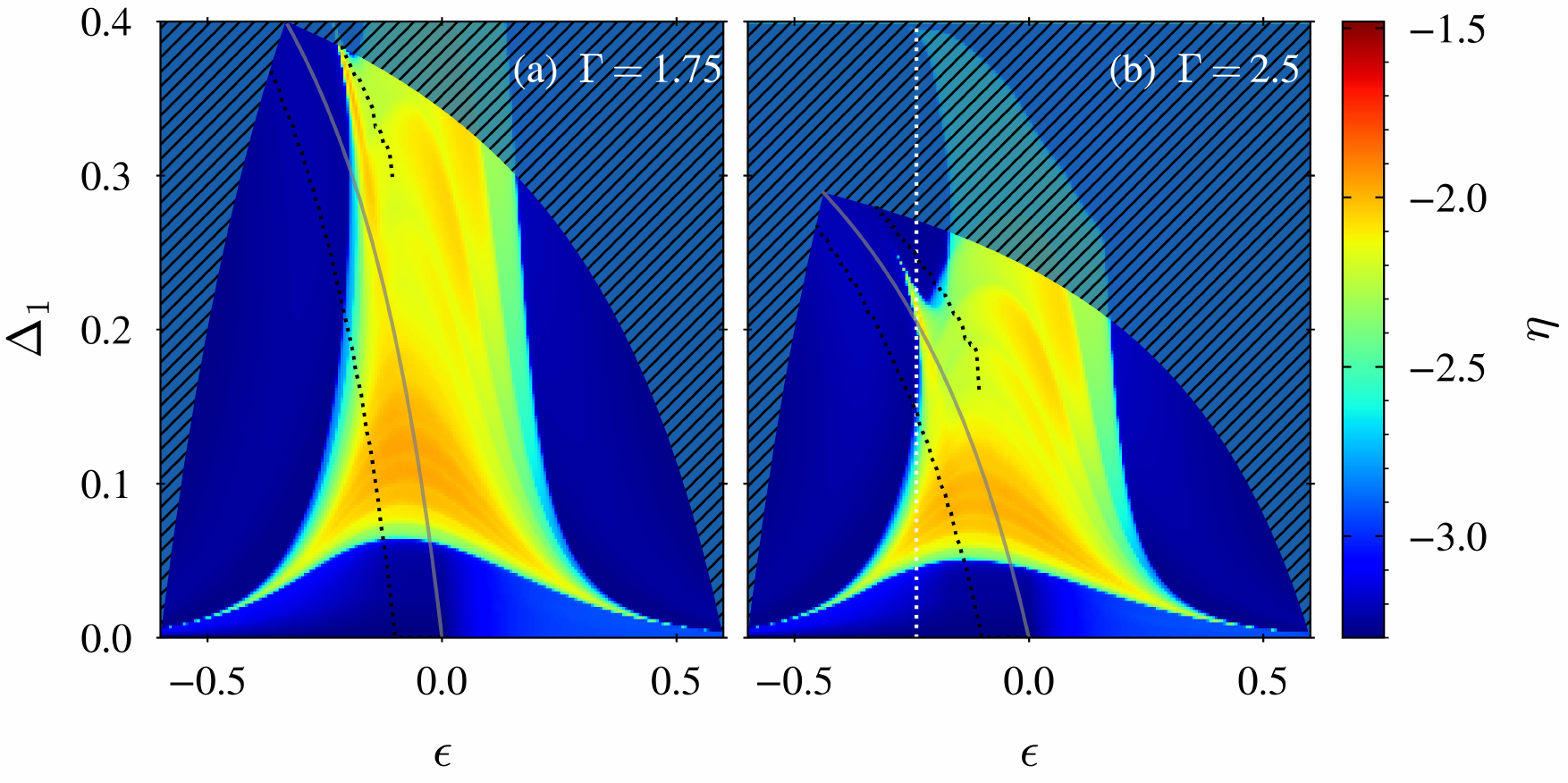}
 \caption{Same quantities as in Fig.~\ref{fig:phase diagrams eta}, but for the case of a chain with nearest-neighbor couplings only. The modulation strength ratio $\Gamma$ is (a) $1.75$ and (b) $2.5$. A clear RLT along the white dotted line as $\Delta_1$ increases is visible in panel (b).
 In both panels, the number of dimers $\mathcal{N}=1000$.}
 \label{fig:n.n. phase diag}
\end{figure}

%%%%%%%%%%%%%%%%%%%%%%%%%%%%%%%%%%%%%%%%%%%%%%%%%%%%%%%%%%%%%%%%%%%%%%%%%%%%%%%%%%%%%%%%
%%%%%%%%%%%%%%%%%%%%%%%%%%%%%%%%%%%%%%%%%%%%%%%%%%%%%%%%%%%%%%%%%%%%%%%%%%%%%%%%%%%%%%%%
%%%%%%%%%%%%%%%%%%%%%%%%%%%%%%%%%%%%%%%%%%%%%%%%%%%%%%%%%%%%%%%%%%%%%%%%%%%%%%%%%%%%%%%%
To make a comparative study of the effects of all-to-all dipolar couplings, we consider in this Appendix the nearest-neighbor approximation of Hamiltonian \eqref{eq:H}.
Since long-range couplings modify the eigenspectrum of the Hamiltonian, we expect the RLT in the nearest-neighbor approximation to take place in a slightly different parameter region.
To precisely find this region, we proceed in a similar way as in the main text and compute in Fig.~\ref{fig:n.n. phase diag} the quantity $\eta$ defined in Eq.~\eqref{eq:eta} in the $(\epsilon,\Delta_1)$ plane.

Figure~\ref{fig:n.n. phase diag}(a) presents the case of a quasiperiodic modulation strength ratio $\Gamma=1.75$, the value used in the main text when considering all-to-all interactions.
While such a localization phase diagram is very similar to the one observed with all-to-all interactions [cf. Fig.~\ref{fig:phase diagrams eta}(b)], some differences are visible.
Notably, within the nearest-neighbor approximation, the first intermediate phase for $\epsilon\neq0$ is smaller, presents a smaller value of $\eta$, and vanishes for large dimerizations.
Moreover, the second intermediate phase is pushed at larger values of $\Delta_1$, so that the RLT appearing for a ratio $\Gamma=1.75$ does not anymore fall inside the regime of parameters constrained by our neglection of multipolar terms [see Eq.~\eqref{eq:dipolar constraint}].
To remedy this issue, we increase $\Gamma$ to $2.5$ in Fig.~\ref{fig:n.n. phase diag}(b), since, as discussed in the main text, it allows to relocate the RLT at smaller values of modulation strength [cf. Fig.~\ref{fig:phase diagrams eta}(b)-(c)].
A clear RLT is now visible along the white dotted line at $\epsilon=-0.24$ as $\Delta_1$ increases.
Interestingly, the simplistic analytical estimate \eqref{eq:analytical curve} based on the local dimerization along the chain discussed in the main text also qualitatively predicts a region where the RLT occurs.
Notice that a second RLT may be observed in the vicinity of $\Delta_1=0.4$. However, such a value of $\Delta_1$ lies beyond the region in which our dipolar treatment holds.
We note that the quantity $\eta$ within the second intermediate phase is here larger than when considering all-to-all interactions.

\begin{figure}[t]
 \includegraphics[width=.5\linewidth]{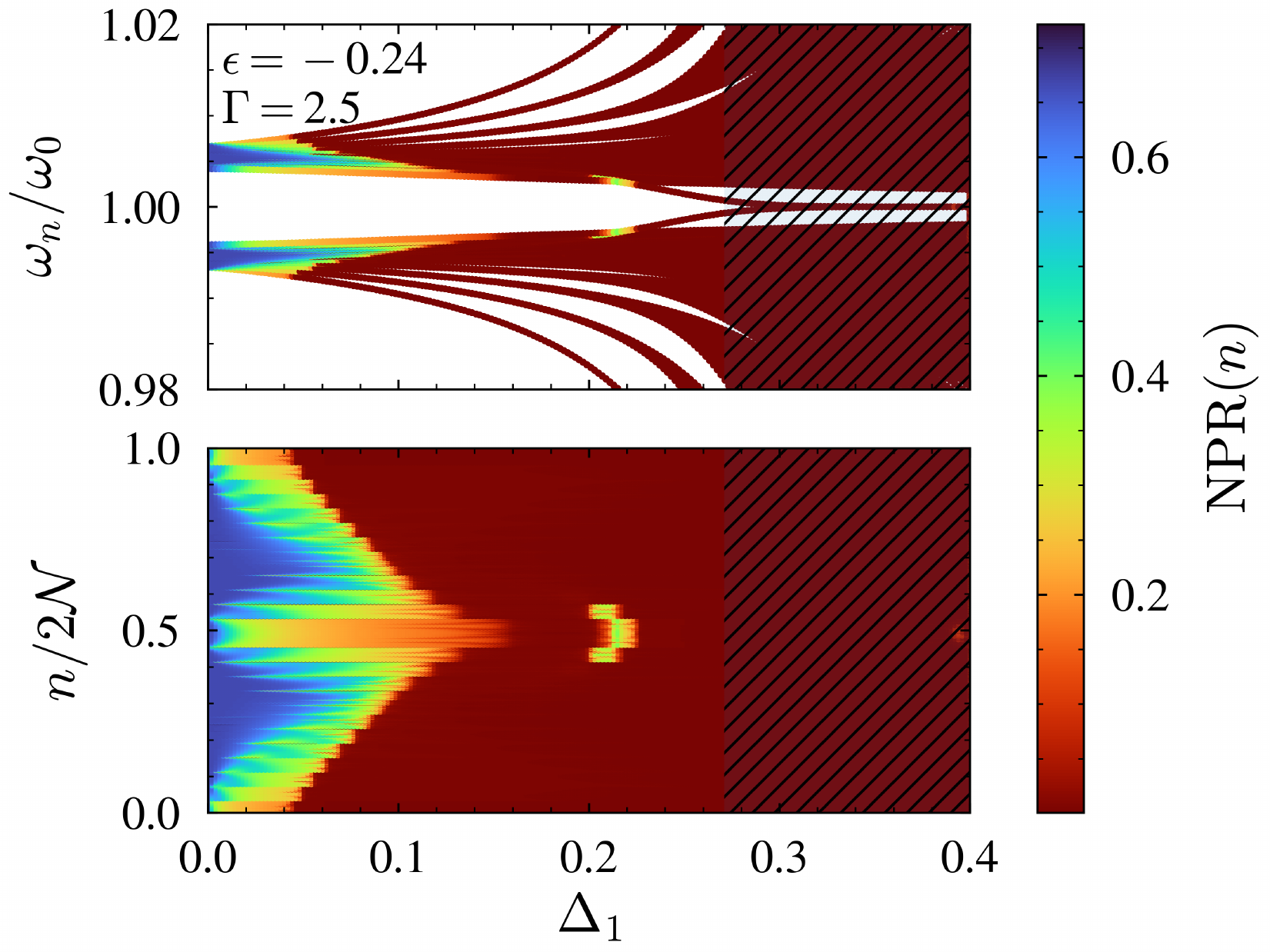}
 \caption{Same quantities as in Fig.~\ref{fig:spectrum vs Delta1}, but for the case of a chain with nearest-neighbor couplings only. A very distinct RLT is visible around $\Delta_1=0.22$ for a large portion of the eigenstates centered in the middle of the spectrum. In the figure, the number of dimers $\mathcal{N}=1000$.}
 \label{fig:n.n. spectrum}
\end{figure}

To better explore the RLT, we display in Fig.~\ref{fig:n.n. spectrum} the counterpart of Fig.~\ref{fig:spectrum vs Delta1} of the main text, namely, the NPR of the eigenspectrum as a function of the quasiperiodic strength $\Delta_1$.
As in Fig.~\ref{fig:n.n. phase diag}, the hatched area corresponds to the parameter region contravening the dipolar constraint \eqref{eq:dipolar constraint}.
Looking at the bandstructure in the upper panel, we notice that from the chiral symmetry of the nearest-neighbor chain we recover a mirror symmetry of the eigenfrequencies around $\omega_0$.
Such a mirror symmetry between the two bands is found to be important for the RLT, as it leads states from both bands, centered in the middle of the spectrum, to enter into a second intermediate phase.
This implies that twice the number of eigenstates is affected by the RLT as compared with the all-to-all dipolar coupling case discussed in the main text, where only states from the lower band are concerned.

Interestingly, we have already seen in the main text that the RLT appears precisely when a miniband enters into the bandgap.
In the nearest-neighbor approximation, once the two symmetric minibands enter in the bandgap, they merge and form a mid-gap miniband of localized states.
However, such an exotic mid-gap miniband of relocalized states precisely forms when the modulation strength is strong enough so that multipolar effects could have an importance in the system, which prevents us to draw any reasonable conclusion.
We verified nevertheless that the eigenstates after the second intermediate phase are localized in the bulk of the chain, and not at its ends.
Moreover, we also verified that the exact same phenomena happens for a nearest-neighbor ring chain of dipoles, i.e., when considering periodic boundary conditions, so that neither such mid-gap states nor the RLT originate from boundary effects.

To conclude this Appendix, these results on the nearest-neighbor approximation demonstrate that in our system, the long-range coupling is globally detrimental to the observation of an RLT, as it strongly reduces the fraction of the concerned eigenspectrum.
This is in accordance with the findings of the authors of Ref.~\cite{Wang2023}, which unveiled in their model that the RLT fades out as the next-to-nearest neighbor coupling is increased, although the way the transition is altered is different in our case.
The effect of chiral symmetry in our model also resonates with the findings of the authors of Ref.~\cite{Chang2025}, that is, the fact that chirally-preserving long-range coupling does not compete against the RLT in their model.
\newpage

%%%%%%%%%%%%%%%%%%%%%%%%%%%%%%%%%%%%%%%%%%%%%%%%%%%%%%%%%%%%%%%%%%%%%%%%%%%%%%%%%%%%%%%%
%%%%%%%%%%%%%%%%%%%%%%%%%%%%%%%%%%%%%%%%%%%%%%%%%%%%%%%%%%%%%%%%%%%%%%%%%%%%%%%%%%%%%%%%
%%%%%%%%%%%%%%%%%%%%%%%%%%%%%%%%%%%%%%%%%%%%%%%%%%%%%%%%%%%%%%%%%%%%%%%%%%%%%%%%%%%%%%%%
\bibliography{Bibliography}
%%%%%%%%%%%%%%%%%%%%%%%%%%%%%%%%%%%%%%%%%%%%%%%%%%%%%%%%%%%%%%%%%%%%%%%%%%%%%%%%%%%%%%%%
%%%%%%%%%%%%%%%%%%%%%%%%%%%%%%%%%%%%%%%%%%%%%%%%%%%%%%%%%%%%%%%%%%%%%%%%%%%%%%%%%%%%%%%%
%%%%%%%%%%%%%%%%%%%%%%%%%%%%%%%%%%%%%%%%%%%%%%%%%%%%%%%%%%%%%%%%%%%%%%%%%%%%%%%%%%%%%%%%

\end{document}